\documentclass[prl,twocolumn,superscriptaddress,floatfix]{revtex4}
\usepackage{CJK}
\usepackage{graphicx}
\usepackage{epstopdf}
\usepackage{color}
\usepackage{ulem}
\usepackage{gensymb}

\begin{document}
\begin{CJK*}{}{}

\title{Robust odd-parity superconductivity in the doped topological insulator Nb$_x$Bi$_2$Se$_3$}
\author{M.~P.~Smylie}
\affiliation{Materials Science Division, Argonne National Laboratory, Argonne, IL 60439}
\affiliation{Department of Physics, University of Notre Dame, Notre Dame, IN 46556}
\author{K.~Willa}
\affiliation{Materials Science Division, Argonne National Laboratory, Argonne, IL 60439}
\author{H.~Claus}
\affiliation{Materials Science Division, Argonne National Laboratory, Argonne, IL 60439}
\author{A.~Snezhko}
\affiliation{Materials Science Division, Argonne National Laboratory, Argonne, IL 60439}
\author{I.~Martin}
\affiliation{Materials Science Division, Argonne National Laboratory, Argonne, IL 60439}
\author{W.-K.~Kwok}
\affiliation{Materials Science Division, Argonne National Laboratory, Argonne, IL 60439}
\author{Y.~Qiu}
\affiliation{Department of Physics, Missouri University of Science and Technology, Rolla, MO 65409}
\author{Y.~S.~Hor}
\affiliation{Department of Physics, Missouri University of Science and Technology, Rolla, MO 65409}
\author{E.~Bokari}
\affiliation{Department of Physics, Western Michigan University, Kalamazoo, MI 49008}
\author{P.~Niraula}
\affiliation{Department of Physics, Western Michigan University, Kalamazoo, MI 49008}
\author{A.~Kayani}
\affiliation{Department of Physics, Western Michigan University, Kalamazoo, MI 49008}
\author{V.~Mishra}
\affiliation{Computer Science and Mathematics Division, Oak Ridge National Laboratory, Oak Ridge, TN 37831}
\author{U.~Welp}
\affiliation{Materials Science Division, Argonne National Laboratory, Argonne, IL 60439}

\date{\today}

\begin{abstract}
We present resistivity and magnetization measurements on proton-irradiated crystals demonstrating that the superconducting state in the doped topological superconductor Nb$_x$Bi$_2$Se$_3$ (x = 0.25) is surprisingly robust against disorder-induced electron scattering.
The superconducting transition temperature $T_c$ decreases without indication of saturation with increasing defect concentration, and the corresponding scattering rates far surpass expectations based on conventional theory.
The low-temperature variation of the London penetration depth $\Delta\lambda(T)$ follows a power law ($\Delta\lambda(T)\sim T^2$) indicating the presence of symmetry-protected point nodes.  
Our results are consistent with the proposed robust nematic $E_u$ pairing state in this material.
\end{abstract}

\maketitle
\end{CJK*}

Topological superconductors have attracted considerable interest \cite{Qi-Zhang-RevModPhys-TI-SC-review,Ando-Fu-AnnualReview-TCI-and-TSC-review,Sato-Ando-arXiv-TSC-review,Sasaki-Mizushima-PhysicaC-SC-doped-TIs,Fu-NatPhys-NewsAndViews-Bi2Se3-SC,Matano-Ando-NatPhys-Knight-Shift-CBS} since they host gapless surface quasi-particle excitations in the form of Majorana fermions.
The non-Abelian braiding properties of Majorana fermions constitute the basis for novel approaches to fault-tolerant quantum computing \cite{Wilczek-NatPhys-Majoranas,Beenakker-AnnRevCMatt-Majoranas-in-TSCs}.
The synthesis of topological superconductors is being pursued along two lines: proximity induced at the interface between conventional superconductors and certain semiconductors with large spin-orbit coupling \cite{Beenakker-AnnRevCMatt-Majoranas-in-TSCs}, or as bulk material obtained by doping topological insulators, for instance Sn$_{1-x}$In$_x$Te \cite{Erickson-Geballe-Fisher-PRB-first-modern-TIT-study,Novak-Ando-PRB-Unusual-gap-in-TIT,Zhong-Gu-PRB-optimize-Tc-in-TIT} and M$_x$Bi$_2$Se$_3$ (M=Cu, Sr, Nb) \cite{Hor-Cava-CBS-discovery,Liu-JACS-SBS-discovery,Qiu-Hor-arXiv-NBS-discovery}.

The emergence of topological superconductivity is determined by the symmetries and dimensionality of the material.
In centro-symmetric and time-reversal invariant superconductors with complete gap \cite{Fu-Berg-PRL-TSC-and-CBS-model,Chiu-Schnyder-RevModPhys-Classification-of-topological-materials} or with nodal gap \cite{Schnyder-Brydon-JPhysConMat-TopNodalSC,Chiu-Schnyder-RevModPhys-Classification-of-topological-materials,Sato-Fujimoto-PRL-Majoranas-in-nodal-TSCs}, the superconducting state will have non-trivial topological characteristics if superconducting pairing has odd-parity, $\Delta$(-\textbf{k}) = -$\Delta$(\textbf{k}), and if the Fermi surface contains an odd number of time-reversal invariant momenta, \textbf{k} = -\textbf{k} + \textbf{G} with \textbf{G} a reciprocal lattice vector.
For weak spin-orbit coupling (i.e., spin is a good quantum number), odd-parity pairing corresponds to spin-triplet pairing.
Odd-parity pairing has been observed in the B-phase of superfluid $^3$He \cite{Vollhardt-He3-superfluidity} and is thought to be realized in several strongly correlated electron systems such as UPt$_3$ or UBe$_{13}$ \cite{Joynt-Taillefer-RevModPhys-UPt3,Gross-Hirschfeld-ZPhysB-spin-triplet-and-lambda} as well as Sr$_2$RuO$_4$ \cite{Mackenzie-Maeno-RevModPhys-spin-triplet-in-SRO}.
In contrast, conventional $s$-wave superconductors are not topological and do not support the Majorana surface mode.

An important unsettled question regarding the realization of topological superconductivity relates to its robustness against disorder in the material.
The effect of electron scattering due to impurities and defects on the superconducting state crucially depends on the structure of the superconducting gap.
Whereas an isotropic fully gapped $s$-wave state is robust against potential scattering due to nonmagnetic impurities \cite{Anderson-JPhysChemSol-Andersons-Theory,Abrikosov-Gorkov-theory}, unconventional superconductors are rather sensitive to disorder \cite{Balatsky-RevModPhys-Nodal-SC-weak-against-disorder}; therefore, one may have expected that topological superconductivity could only be achieved in extremely clean samples.
However, recent theoretical considerations \cite{Nagai-Ota-PRB-TSC-may-be-robust,Nagai-PRB-TSC-is-robust-in-CBS-theory,Michaeli-Fu-PRL-Odd-parity-robustness} show odd-parity topological superconductivity with strong spin-orbit coupling may in fact be robust against disorder.
Here, we present a study of the evolution of $T_c$, of the low-temperature London penetration depth $\lambda$, and of the resistivity of the candidate topological superconductor Nb$_x$Bi$_2$Se$_3$ with increasing disorder as introduced by proton irradiation.
In the covered temperature range ($T/T_c \geq$ 0.12) the temperature variation of $\lambda(T)$ of the pristine samples as well as of all irradiated crystals is quadratic, $\Delta\lambda(T)\sim T^2$, indicative of symmetry-protected point nodes.
$T_c$ is suppressed with increasing proton dose in all crystals, with no trend towards saturation at high doses.
Concurrently, the residual resistivity, $\rho_0$, increases strongly.
Within the conventional Abrikosov-Gor`kov theory \cite{Abrikosov-Gorkov-theory}, such increase of $\rho_0$ would induce two orders of magnitude stronger suppression of $T_c$, which suggests that the superconducting state is indeed robust against impurity scattering, contrary to more conventional nodal superconductors.

High-quality crystals of Nb$_x$Bi$_2$Se$_3$ with high superconducting volume fractions were grown by the same method used in Ref.~\onlinecite{Qiu-Hor-arXiv-NBS-discovery}, and show high superconducting volume fractions.
Nb$_x$Bi$_2$Se$_3$ has the same trigonal space group $R\bar{3}m$ as the parent compound Bi$_2$Se$_3$, with slightly expanded $c$ axis to accommodate the Nb ion interstitially between adjacent Bi$_2$Se$_3$ quintuple layers (see Fig. 1).
All samples were repeatedly irradiated along the $c$ axis with 5 MeV protons using the tandem Van de Graaff accelerator at Western Michigan University.
The proton beam was passed through a gold foil to ensure homogeneous irradiation, and the sample was cooled to $\sim$-10\degree C during irradiation.
TRIM simulations \cite{SRIM-reference-book} for our irradiation geometry show that defect generation is uniform through the thickness of the samples.
Irradiation with MeV-protons creates a distribution of defects including point defects in the form of interstitial-vacancy pairs as well as collision cascades and clusters \cite{Kirk-Yan-defect-generation,LeiFang-Welp-Kwok-PRB-defect-generation-in-Ba122P,Civale-PRL-proton-damage-in-YBCO}.

\begin{figure}
\includegraphics[width=1\columnwidth]{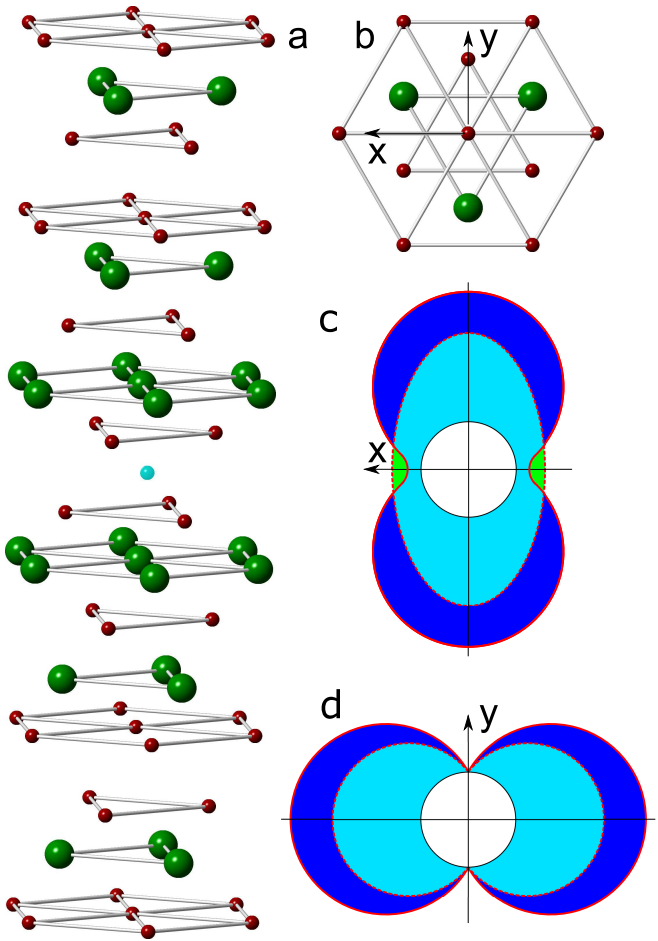}
\caption{
a) The crystal structure of Nb$_x$Bi$_2$Se$_3$ derived from an ABC stacking of hexagonal sheets of Bi (green) and Se (red) atoms.
The Nb ions (blue) sit in the van der Waals gap between quintuple layers of Bi$_2$Se$_3$ [21].
b) View down the $c$ axis, the $y$ axis is chosen to lie in the mirror plane.
c) and d) Schematic presentations of the effect of defect scattering on an $s$-wave gap with symmetry-protected point nodes (d) and a gap with deep minima (c).
Dark and light blue represent the gap amplitude before and after introduction of scattering, respectively.
}
\label{figStructure}
\end{figure}

We performed $ac$-susceptibility and London penetration depth measurements using the tunnel-diode oscillator (TDO) technique \cite{Prozorov-Giannetta-SST-TDO-reference} employing a custom-built TDO operating at 14.5 MHz.
Here, the change in the resonator frequency $\Delta f(T)$ is proportional to the change of the London penetration depth $\Delta\lambda(T)$ such that $\Delta f(T)/\Delta f_0 = G\Delta\lambda(T) / \lambda_0$, where G is a calibration factor, $\Delta f_0$ is the total frequency change occuring between the lowest temperature and $T_c$, and $\lambda_0 = \lambda(T = 0)$.
All crystals measured here showed $T_c \approx$ 3.4 K in the pristine state with minimal sample-to-sample variation.

\begin{figure}
\includegraphics[width=1\columnwidth]{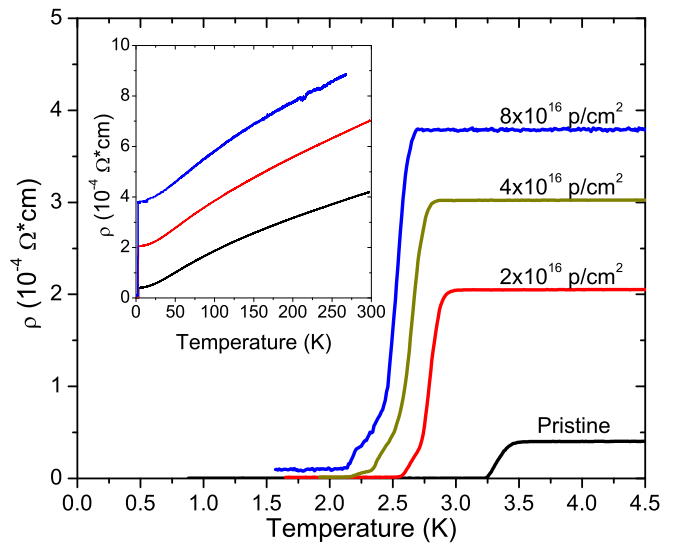}
\caption{
Low-temperature resistivity of a single crystal of Nb$_x$Bi$_2$Se$_3$ showing suppression of $T_c$ and increase in the residual resistivity $\rho_0$ following multiple irradiations.
The inset shows the resistivity up to room temperature with little change in curvature following repeated doses.
}
\label{figTransport}
\end{figure}

In Fig. 2, the temperature dependence of the resistivity for multiple irradiation levels measured up to room temperature is shown in the inset.
The irradiation does not significantly affect the curvature of $\rho$ vs $T$, but instead offsets the curves, consistent with an increase in residual resistivity $\rho_0$.
As the cumulative proton dose is increased, the transition temperature is clearly suppressed and the residual resistivity $\rho_0$, taken as the effectively temperature-independent value of the resistivity just above the transition onset, increases strongly.
For all doses, the transitions remain reasonably sharp, indicating single-phase behavior throughout.
The temperature dependent normalized magnetic susceptibility as determined from the TDO frequency shift of one sample is shown in Fig. 3 for multiple irradiation doses.
The superconducting transition temperature $T_c$ is clearly suppressed with each dose.
Nevertheless, the transitions remain sharp even at the highest cumulative irradiation dose.
No secondary transitions from possible superconducting contaminants Nb or NbSe$_2$ were observed at higher temperatures.

\begin{figure}
\includegraphics[width=1\columnwidth]{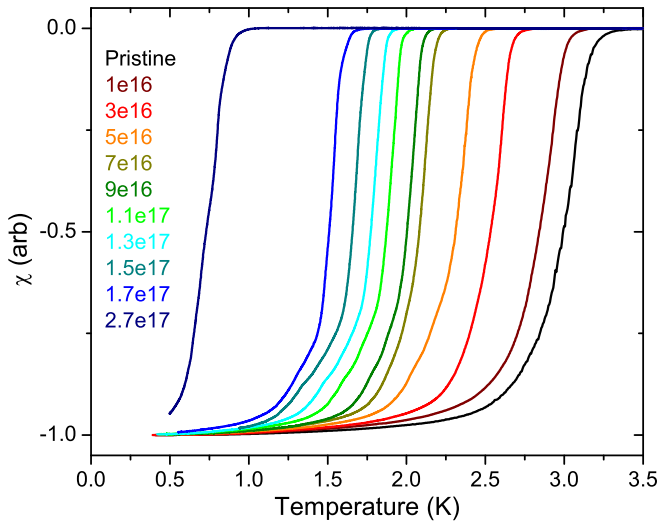}
\caption{
Normalized magnetic susceptibility of a single crystal of Nb$_x$Bi$_2$Se$_3$ as a function of temperature for various values of cumulative p-irradiation dose.
The transition temperature $T_c$ is clearly suppressed with each dose given in p/cm$^2$. 
}
\label{figTDOdose}
\end{figure}

Fig. 4 shows the low-temperature variation of the penetration depth $\Delta\lambda(T)$ of a Nb$_x$Bi$_2$Se$_3$ crystal irradiated to several cumulative doses versus reduced temperature squared, $(T/T_c)^2$.
These data reveal that in the measured temperature range the penetration depth has a quadratic temperature dependence, $\Delta\lambda(T) \sim T^2$ for all doses of irradiation.
The low-temperature variation of the London penetration depth is determined by the distribution and scattering of thermally activated quasi-particles on the Fermi surface.
For an isotropic $s$-wave superconductor, $\Delta\lambda(T)$ at sufficiently low temperatures follows an exponential variation, $\Delta\lambda(T)/\lambda_0 \approx \sqrt{\pi\Delta_0 / 2T}$exp$(-\Delta_0 / T)$
where $\Delta_0$ is the zero temperature value of the energy gap.
Nodes in the gap, however, will induce enhanced thermal excitation of low-lying quasi-particles, resulting in a power-law variation, $\Delta\lambda \sim T^n$, with the exponent depending on the type of node and on electron scattering.
In particular, a quadratic temperature dependence is expected in a clean material with linear quasiparticle dispersion around point nodes in the superconducting gap \cite{Gross-Hirschfeld-ZPhysB-spin-triplet-and-lambda}.
The observation \cite{Lawson-LuLi-dHvA-on-NBS-Fermi-surfaces} of quantum oscillations in Nb$_x$Bi$_2$Se$_3$ crystals similar to those used here shows that the unirradiated samples are fairly clean.
Hence the quadratic temperature dependence of $\lambda$ is indicative of point nodes \cite{Smylie-TDO-on-NBS}. 

\begin{figure}
\includegraphics[width=1\columnwidth]{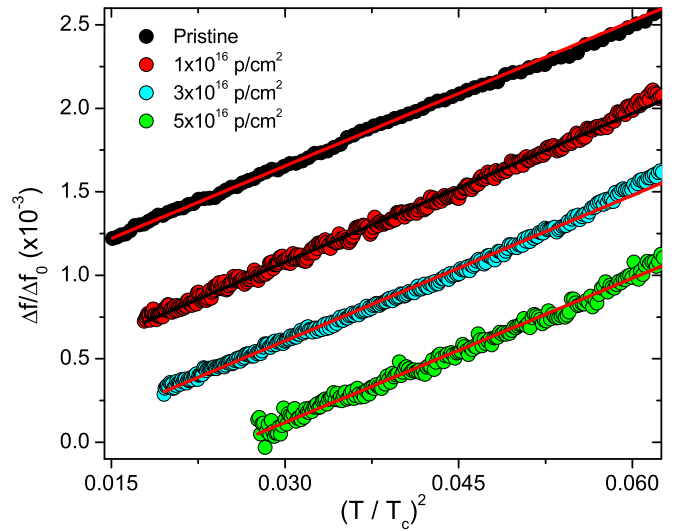}
\caption{
Low-temperature variation of the London penetration depth $\Delta\lambda(T)$ in a single crystal of Nb$_x$Bi$_2$Se$_3$ for multiple values of cumulative irradiation dose vs reduced temperature squared $(T/T_c)^2$.
The linear fits (red, black lines) indicate quadratic behavior.
As the dose increases, the temperature dependence remains quadratic, indicative of point nodes in the superconducting gap.
Data are off-set vertically for clarity of presentation.
}
\label{figTDOlowT}
\end{figure}

As shown in Fig. 4, the temperature dependence of $\lambda$ remains quadratic with increasing disorder.
This finding is consistent with a theoretical analysis of the effect of impurity scattering on the gap structure of $p$-wave superconductors \cite{Gross-Hirschfeld-ZPhysB-spin-triplet-and-lambda}.
For the axial $p$-wave gap (two point nodes) impurity scattering rates below a critical value do not affect the $T^2$-dependence.
In contrast, the linear temperature dependence of $\lambda$ expected for the polar $p$-wave gap (equatorial line node) is expected to be strongly affected by impurity scattering.

A $T^2$-dependence of $\lambda$ could also arise in an anisotropic $s$-wave gap with deep gap minima such that the minimum gap is significantly smaller than the measurement temperature.
However, potential scattering will make an anisotropic $s$-wave gap more isotropic implying an increase in the minimum gap value with increasing scattering \cite{Borkowski-Hirschfeld-PRB-Hirschfeld-scattering,Fehrenbacher-Norman-PRB-scattering-in-anisotropic-materials} (see Fig. 1c) thereby altering the low-temperature variation of the penetration depth.
In contrast, as symmetry-imposed nodes in the gap cannot be removed by electron scattering, the gap amplitude decreases rapidly with increasing scattering rate while the overall gap structure remains unchanged as indicated in Fig. 1d.
Therefore, the persistent $T^2$-variation in the data in Fig. 4 rules out an anisotropic $s$-wave gap with deep minima, and is further support for an unconventional superconducting gap in Nb$_x$Bi$_2$Se$_3$.

Theoretical analysis of Bi$_2$Se$_3$-based superconductors \cite{Fu-Berg-PRL-TSC-and-CBS-model,Fu-PRB-explaining-Knight-shift-in-CBS,Venderbos-Kozii-Fu-PRB-Two-component-order-parameters-in-Bi2Se3s} shows that strong spin-orbit coupling can induce unconventional pairing symmetries in time-reversal symmetric systems, even if the pairing is mediated by conventional electron-phonon coupling.
In particular, in a two-orbital model with short-range pairing interactions four pairing states that transform according to the four irreducible representations of the $D_{3d}$ crystal point group of Nb$_x$Bi$_2$Se$_3$ were identified.
One is the fully symmetric conventional $s$-wave state, whereas the other three have odd-parity pairing.
Among the latter, the state that corresponds to the two-dimensional representation $E_u$ has attracted considerable attention as it allows for a nematic state that would account for the surprising two-fold symmetry that emerges in several quantities below $T_c$ \cite{Fu-PRB-explaining-Knight-shift-in-CBS}, i.e., the Knight shift and specific heat in Cu$_x$Bi$_2$Se$_3$ \cite{Matano-Ando-NatPhys-Knight-Shift-CBS,Yonezawa-Ando-NatPhys-Rotational-breaking-via-calorimetry-in-CBS}, magneto-transport \cite{Pan-deVisser-SciRep-Rotational-breaking-via-transport-in-SBS,Du-Gu-HHWen-arXiv-Corbino-geometry-transport-rotational-breaking-in-SBS} in Sr$_x$Bi$_2$Se$_3$, and magnetic torque \cite{Asaba-Lawson-LuLi-PRX-Rotational-breaking-via-torque-magnetometry-in-NBS} in Nb$_x$Bi$_2$Se$_3$.
The gap structure of the $E_u$-state depends on the orientation of the nematic director \textbf{n} (see Fig. 1); for \textbf{n} along an $x$-axis (perpendicular to the mirror plane) the $\Delta_{4x}$ state is realized with two symmetry-protected point nodes along \textbf{k}$_y$, whereas for \textbf{n} along a $y$-axis (parallel to the mirror plane) the $\Delta_{4y}$ state emerges with gap minima along \textbf{k}$_x$ \cite{Fu-PRB-explaining-Knight-shift-in-CBS}.
A detailed study analogous to \cite{Borkowski-Hirschfeld-PRB-Hirschfeld-scattering,Fehrenbacher-Norman-PRB-scattering-in-anisotropic-materials} of the response of the gap minima in the $\Delta_{4y}$ state to electron scattering has not been discussed yet in the literature to our knowledge.
However, Fig. 4 shows that at reduced temperatures as low as 0.12 there is no indication of deviation from the $T^2$-dependence of $\lambda$ which would imply a very large ratio of maximum and minimum gap in a possible $\Delta_{4y}$ state of more than 10.
Thus, while it is difficult to rule out $\Delta_{4y}$ completely, our results point towards the $\Delta_{4x}$ state as the superconducting ground state of Nb$_x$Bi$_2$Se$_3$.

Fig. 5 summarizes the evolution of $T_c$ with increasing proton irradiation dose as determined from resistivity and $ac$-susceptibility measurements.
For the transport measurement samples, the increase of the residual resistivity, $\Delta\rho_0$, is directly obtained (see Fig. 2), whereas for the TDO samples, the $\Delta\rho_0$ values corresponding to a given p-dose are inferred from a fit of $\Delta\rho_0$ versus proton dose data obtained from the transport samples.
Although there is some scatter in the data, the $T_c$ values of all Nb$_x$Bi$_2$Se$_3$ samples follow a smooth trend towards $T_c$ = 0 with increasing dose without any indication of saturation.
The lack of saturation reinforces a model of odd-parity superconductivity in Nb$_x$Bi$_2$Se$_3$.

\begin{figure}
\includegraphics[width=1\columnwidth]{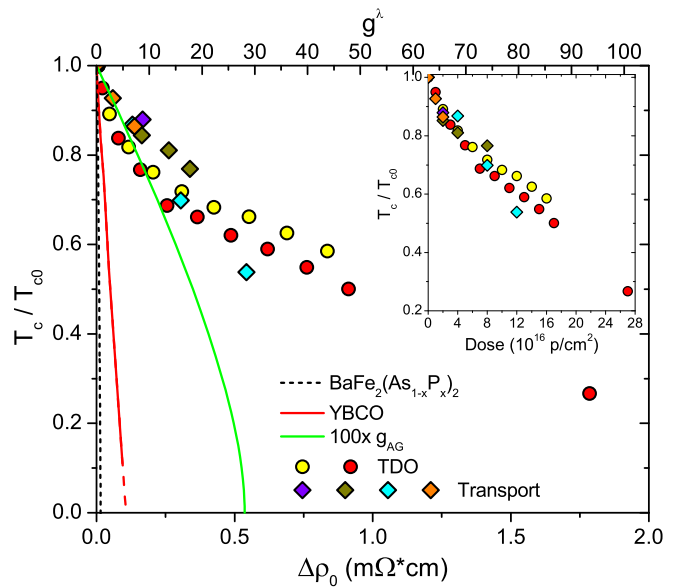}
\caption{
Evolution of $T_c$ with proton irradiation for several crystals of Nb$_x$Bi$_2$Se$_3$, as measured via transport (diamonds) and magnetic susceptibility (circles).
The mustard diamonds and red circles are derived from the samples shown in Fig. 2 and 3, respectively.
The inset shows $T_c/T_{c0}$ versus proton dose, whereas the main panel displays the same data as function of increase in residual resistivity, $\Delta\rho_0$ (lower x-axis) and normalized scattering rate, $g^\lambda$ (top x-axis).
For comparison, the $T_c/T_{c0}$ data versus $g^\lambda$ of p-irradiated BaFe$_2$(As$_{1-x}$P$_x$)$_2$ (dashed line), He$^+$-irradiated YBCO (red line), and the Abrikosov-Gor`kov prediction for $T_c/T_{c0}$ versus $g$ on a 100x expanded scale (green line) are included. 
}
\label{figG}
\end{figure}

Further analysis of the data in Fig. 5 is based on the Abrikosov-Gor`kov (AG) theory of pair-breaking scattering \cite{Abrikosov-Gorkov-theory,Openov-PRB-Disorder-scattering-beyond-AG-theory}.
For magnetic scattering in isotropic $s$-wave superconductors, or for potential scattering in superconductors with an anisotropic gap, the suppression of $T_c$ is given as ln $(T_c/T_{c0}) = \chi[\Psi(1/2) - \Psi(gT_{c0} / 2T_c)]$.
Here, $\Psi$ is the digamma function, $\chi$ is a measure of the gap anisotropy, and $g = \hbar / 2\pi k_B T_{c0}\tau$ is the normalized scattering rate with $\tau$ the pairbreaking scattering time.
For nonmagnetic defects, $\tau$ corresponds to the potential scattering time and for magnetic impurities to half the spin-flip scattering time.
Since for odd-parity pairing, the Fermi surface average of $\Delta$(\textbf{k}) is zero, $\chi=1$, and $T_c$ is suppressed to zero at a critical value $g_c \approx$ 0.28.
Linking the scattering rate to measurable quantities such as the increase in resistivity requires detailed information on the electronic band structure, transport and particle lifetimes, and the scattering potential.
For instance, it has previously been observed that in multi-band superconductors the suppression of $T_c$ with disorder depends sensitively on the balance between inter and intra-band scattering rates \cite{Prozorov-Hirschfeld-PRX-Multiband-scattering}.
As many of the microscopic parameters of Nb$_x$Bi$_2$Se$_3$ are currently unavailable we relate the measured increase in resistivity to the scattering rate using a simple single-band Drude model, $\Delta\rho_0 = m^* / (ne^2\tau_i)$ with $m^*$ and $n$ the effective mass and concentration of carriers, respectively, and 1/$\tau_i$ the scattering rate due to the irradiation-induced defects.
Since the enhancement of the residual resistivity is large, we neglect the contribution from pre-existing defects in the total scattering rate.
The parameter $m^* /ne^2$ can be estimated from values of the penetration depth, $\lambda^2 = m^* / \mu_0 ne^2$.
We thus obtain the normalized scattering rate $g$ in terms of the London penetration depth as $g^\lambda = \hbar \Delta\rho_0 / 2 \pi k_B T_{c0} \mu_0 \lambda^2$,
yielding $g^\lambda \approx 0.172~\Delta\rho_0/T_{c0}$, where $\Delta\rho_0$ is expressed in $\mu\Omega\cdot$cm and with a zero-temperature penetration depth of $\sim$ 240 nm \cite{Smylie-TDO-on-NBS}.

The data in Fig. 5 show that the increase in resistivity required to induce a given reduction of $T_c$ is enhanced over predictions based on the AG theory by a very large margin.
In AG-theory, each scattering event giving rise to enhanced resistivity is also pair-breaking.
This implies that in Nb$_x$Bi$_2$Se$_3$ the majority of scattering events do not contribute to pair-breaking.
Also included in Fig. 5 are the $T_c$/T$_{c0}$ vs $g^\lambda$ data on proton-irradiated BaFe$_2$(As$_{1-x}$P$_x$)$_2$ \cite{Smylie-PRB-TDO-on-Ba122P} and on He$^+$-irradiated YBCO \cite{Lang-YBCO-Tc-suppression}.
These materials have sign-changing order parameters--$s_\pm$-gap symmetry with additional accidental line nodes and $d$-wave symmetry, respectively.
Therefore, nonmagnetic potential scattering induces a rapid suppression of $T_c$.
Similar behavior would be expected for Nb$_x$Bi$_2$Se$_3$ due to odd-parity pairing.
Nevertheless, its $T_c$-suppression is in comparison remarkably weak.
The reason for these surprising results lies in the particular electronic structure of Nb$_x$Bi$_2$Se$_3$, which has very strong spin-orbit coupling.
In the relativistic limit of vanishing Dirac mass \cite{Nagai-Ota-PRB-TSC-may-be-robust,Nagai-PRB-TSC-is-robust-in-CBS-theory,Michaeli-Fu-PRL-Odd-parity-robustness}, the emergent chiral symmetry effectively protects against impurity-induced scattering between two pseudo-chiral bands \cite{Nagai-PRB-TSC-is-robust-in-CBS-theory}, if the scattering is non-magnetic and does not discriminate between the pseudo-chiral sectors.
This effectively puts these odd-parity superconductors in the same category with respect to potential impurity scattering as $s$-wave superconductors, protected by the Anderson theorem \cite{Anderson-JPhysChemSol-Andersons-Theory}.
Finite suppression of $T_c$ can result either from magnetic impurities, from disorder that couples differently to the bands, or from combined effect of a finite Dirac mass (that breaks chiral symmetry) and potential scattering.
While irradiation is unlikely to introduce magnetic scattering, both latter mechanisms are most likely present.
Further studies would be necessary to determine their relative importance.

In summary, 5-MeV proton irradiation has been shown to increase electron scattering in the candidate topological superconductor Nb$_x$Bi$_2$Se$_3$.
A substantial increase in electron scattering is required to suppress $T_c$, far larger than anticipated via conventional theory, and the effect does not saturate even at large doses.
The low-temperature variation of the London penetration depth $\Delta\lambda(T)$ remains quadratic in the pristine and disordered states.
Together, these results suggest the presence of symmetry-protected point nodes in Nb$_x$Bi$_2$Se$_3$, further supporting the proposed nematic $E_u$ pairing state.
Owing to the strong spin-orbit locking, these results are the first demonstration of an unconventional superconductor that is robust against nonmagnetic disorder suggesting that topological superconductivity can be realized in rather dirty materials.

\begin{acknowledgments}
TDO and magnetization measurements were supported by the U.S. Department of Energy, Office of Science, Basic Energy Sciences, Materials Sciences and Engineering Division.
MPS thanks ND Energy for supporting his research and professional development through the ND Energy Postdoctoral Fellowship Program.
KW acknowledges support through an Early Postdoc Mobility Fellowship of the Swiss National Science Foundation.
YSH acknowledges support from the National Science Foundation grant number DMR-1255607.
VM was supported by the Laboratory Directed Research and Development Program of Oak Ridge National Laboratory, managed by UT-Battelle, LLC, for the U.S. Department of Energy.
\end{acknowledgments}

\bibliographystyle{apsrev4}

\end{document}